\documentclass[9pt,conference]{IEEEtran}
\usepackage{commath}
\usepackage{booktabs}
\usepackage{placeins}
\usepackage{tabularx}

\usepackage[preprint]{waspaa25}

\usepackage{bm} 

\title{Musical Source Separation Bake-Off: \\
Comparing Objective Metrics with Human Perception}

\name{Noah Jaffe,
      John Ashley Burgoyne}
\address{Institute for Logic, Language, and Computation, University of Amsterdam, The Netherlands}

\begin{document}

\maketitle

\begin{abstract}
Music source separation aims to extract individual sound sources (e.g., vocals, drums, guitar) from a mixed music recording. However, evaluating the quality of separated audio remains challenging, as commonly used metrics like the source-to-distortion ratio (SDR) do not always align with human perception. In this study, we conducted a large-scale listener evaluation on the MUSDB18 test set, collecting approximately 30 ratings per track from seven distinct listener groups. 
We compared several objective energy-ratio metrics, including legacy measures (BSSEval v4, SI-SDR variants), and embedding-based alternatives (Fréchet Audio Distance using CLAP-LAION-music, EnCodec, VGGish, Wave2Vec2, and HuBERT).
While SDR remains the best-performing metric for vocal estimates, our results show that the scale-invariant signal-to-artifacts ratio (SI-SAR) better predicts listener ratings for drums and bass stems. Fréchet Audio Distance (FAD) computed with the CLAP-LAION-music embedding also performs competitively—achieving Kendall’s $\tau$ values of 0.25 for drums and 0.19 for bass—matching or surpassing energy-based metrics for those stems. However, none of the embedding-based metrics, including CLAP, correlate positively with human perception for vocal estimates. These findings highlight the need for stem-specific evaluation strategies and suggest that no single metric reliably reflects perceptual quality across all source types. We release our raw listener ratings to support reproducibility and further research.
\end{abstract}

\section{Introduction}
\label{sec:intro}
Music source separation involves extracting individual
sound sources—called stems—such as vocals, drums, and guitar, from a
mixed music recording. Practical applications include karaoke or
a reprocessing step for many music information retrieval (MIR) tasks. Music source separation systems are typically evaluated using objective metrics such as the source-to-distortion ratio (SDR), sources-to-artifacts ratio (SAR), and source-to-interferences ratio (SIR) \cite{Vincent2006}. Despite their widespread use, these energy-ratio metrics often correlate poorly with human perception, leading to discrepancies between objective evaluations and listener judgments \cite{Cano2016, Emiya2011, rumboldCORRELATIONSOBJECTIVESUBJECTIVE}. Recent work attempted to address shortcomings by proposing scale-invariant alternatives, such as the scale-invariant signal-to-distortion ratio (SI-SDR) and scale-invariant signal-to-artifacts ratio (SI-SAR) \cite{LeRoux2019}. However, even with these improvements, a significant gap remains between numerical scores and listener perception.

In evaluating the performance of source separation systems, a set of original, un-mixed sources – known as stems
or ground truths – are needed for comparison with the estimated stems. A state-of-the-art musical source separation system for pop music typically produces four estimates: \emph{vocals}, \emph{bass}, \emph{drums}, and \emph{other}—encompassing remaining instruments such as guitar or piano.

In this work, we present a comprehensive large-scale listening study that re-evaluates a range of objective metrics on the MUSDB18 test dataset. Our study includes several separation systems covering both legacy methods (e.g., REP1) and state-of-the-art deep-learning models (e.g., Open-Unmix, HTDemucs-ft, SCNet-large), as well as an oracle method (IRM1). Using a webMUSHRA test \cite{ITU-R2015}, we collected approximately 30 ratings per track from participants with a minimum of two years of musical experience. 

The key contributions of this paper are as follows.
\begin{itemize}
    \item We conduct a robust, large-scale listener evaluation.
    \item We compare legacy metrics (e.g., SDR, SI-SDR) with Fréchet Audio Distance (FAD) computed using different embeddings.
    \item We release the complete raw data from the listening study, promoting transparency and enabling future meta-analyses.
\end{itemize}
\section{Overview of Metrics}
\label{metrics}
\subsection{Energy-ratio metrics}
Musical source separation systems are evaluated using a dataset of \emph{stems}, which are then mixed together to create the final recording.
Objective energy-ratio metrics such as SDR and its variants quantify the extent to which an estimate differs from its ground truth stem. The estimate $ \hat{s} $ is decomposed into its ground-truth content $ s_{\text{target}} $ and residual error $  e_{\text{residual}} $.
\begin{equation}\label{components}
\hat{s}=s_{\text{target}} + e_{\text{residual}}
\end{equation} 
The error term is decomposed further based on its origin and can be used to calculate metrics such as the \textit{sources to artifacts ratio} (SAR) and \textit{signal to interference ratio} (SIR).

For traditional SDR, SIR, and SAR in BSSEval v4,\footnote{https://github.com/sigsep/sigsep-mus-eval}
\begin{equation}\label{errorcomponents}
e_{\text{residual}} = e_{\text{spatial}} + e_{\text{interference}} + e_\text{artifact}
\end{equation}
These components form the basis for the log ratio for source-to-distortion ratio (SDR) \cite{Vincent2006}:
\begin{equation}\label{sdr}
\text{SDR} = 10 \log_{10} \frac{\norm{s_{\text{target}}}^{2}} {\norm{e_{\text{spatial}} + e_{\text{interference}} + e_{\text{artifact}}}^{2}}
\end{equation}

As expressed in \eqref{sdr}, SDR compares the energy originating from the original, un-mixed stem with noise sources present in the extracted estimate \cite{Vincent2006}.

In Le Roux et al.’s SI-SDR paper, the authors demonstrate that standard SDR is highly sensitive to gain changes: rescaling an estimate by any constant factor (without altering its perceptual quality) still yields different SDR scores \cite{LeRoux2019}. To avoid this, they introduced scale-invariant variants (SI-SDR, SI-SIR, SI-SAR) that normalize out signal energy differences so the metric reflects only perceptual fidelity rather than arbitrary amplitude scaling. SI-SDR also addresses shortcomings in BSSEval’s time-invariant 512-tap decomposition filter, which in traditional SDR can mask errors by placing spectral nulls and thus forgive distortions in those bands. These metrics, computed on mono tracks, decompose the error term exclusively into interference and artifact components, as illustrated in \eqref{errorcomponentssisdr} relative to \eqref{errorcomponents}. In particular, SI-SDR introduces an explicit scaling step: the optimal scaling factor, $\alpha$, defined as  
$\alpha = \hat{s}^T s\,/\,\lVert s\rVert^2$
from which the scaled reference is defined as  
$ e_{\mathrm{target}} = \alpha\,s.$   
The estimate is then split as  
  $ \hat{s} = e_{\mathrm{target}} + e_{\mathrm{res}},  $
where \(e_{\mathrm{res}}\) is the residual error, yielding the expanded SI-SDR (and by extension SI-SAR, SI-SIR) formulas as seen in \eqref{si-sdr}, \eqref{sisir}, and \eqref{sisar}.  

\begin{equation}\label{errorcomponentssisdr}
e_{\text{residual}} = e_{\text{interference}} + e_\text{artifact}
\end{equation}

\begin{equation}\label{si-sdr}
\text{SI-SDR} = 10 \log_{10} \frac{\norm{e_{\text{target}}}^{2}} {\norm{e_{\text{interference}} + e_{\text{artifact}}}^{2}}
\end{equation}

\begin{equation}\label{sisir}
\text{SI-SIR} = 10 \log_{10} \frac{\norm{e_{\text{target}}}^{2}} {\norm{e_{\text{interference}}}^{2}}
\end{equation}

\begin{equation}\label{sisar}
\text{SI-SAR} = 10 \log_{10} \frac{\norm{e_{\text{target}}}^{2}} {\norm{e_{\text{artifact}}}^{2}}
\end{equation}
In addition, SI-SIR \eqref{sisir} and SI-SAR \eqref{sisar} are simplified and made orthogonal---original formulas for SAR contained noise error terms in the numerator, which made the metric problematic.

Although SI-SDR has seen some adoption, SDR remains the de facto metric used in music source separation evaluation.

\subsection{Perception-informed metrics}
A first effort to create a listener-informed objective metric was the development of the Perceptual Evaluation methods for Audio Source Separation (\texttt{PEASS}) toolkit in 2010 \cite{PEASS2010}. Implemented in MATLAB, PEASS uses nonlinear neural networks to generate multiple perceptual scores aimed at predicting human judgments of separated audio signals. Although innovative in its attempt to capture complex, perceptually relevant aspects of audio quality, PEASS has not achieved widespread adoption.

The Fréchet Audio Distance (FAD) was originally introduced by Kilgour et al. \cite{kilgourFrechetAudioDistance2019} as a novel deep-learning metric that compares the statistical distributions of audio embeddings, rather than merely relying on an objective energy-based ratio. This approach can theoretically offer a better assessment of perceptual audio quality because it captures subtle, high-level characteristics that energy metrics might overlook. In 2024, Microsoft Research augmented FAD with \texttt{fadtk} \cite{fadtk},\footnote{https://github.com/microsoft/fadtk} which extends its application beyond the original VGGish\cite{vggish} embedding to include alternatives like CLAP-LAION-music\cite{laionclap2023}, EnCodec\cite{defossez2022high}, Wave2Vec2\cite{w2v2}, and HuBERT\cite{HuBERT}. Using an embedding model pretrained on audio more similar to the target material—such as music (CLAP, EnCodec) versus speech (HuBERT, Wave2Vec2)—can lead to more accurate representations of relevant features, potentially improving correlation with human perception and the reliability of quality assessments.

\section{Methodology}
\label{sec:methodology}
\subsection{Listener Study}
To obtain perceptual ground truth for evaluating objective metrics, we conducted a listener study. Participants evaluated every stem from every track in the MUSDB18 test dataset (50 tracks). Our study builds on the work conducted in the 2018 Signal Separation Evaluation Campaign (SiSEC2018), which established standard benchmarks and evaluation procedures for music source separation \cite{sisec18}. For each track, a 10-second fragment was extracted beginning at the start time specified in the SiSEC2018 cutlist\footnote{https://github.com/sigsep/sigsep-mus-cutlist-generator}.

We use estimates spanning a range of expected quality levels produced by several separation systems. Estimates were generated using state-of-the-art models such as Hybrid Transformer Demucs (HT-Demucs) \cite{htdemucs} and Sparse Compression Network (SCNet) \cite{scnet}. In addition, estimates were produced using Open-Unmix \cite{openunmix}, while the SiSEC2018 submissions\footnote{https://zenodo.org/records/1256003} provided estimates for the Ideal Ratio Mask (IRM1)—an oracle method \cite{sisec18}—and the Repeating Pattern Extraction Technique (REPET)\footnote{Only vocals estimates were used from REPET.}—a legacy BSS method \cite{Rafii2013}.

To reduce listener fatigue and ensure high-quality responses, we conducted the evaluation online using the \texttt{webMUSHRA} platform \cite{schoeffler2015mushra} in seven separate batches. In contrast to the official MUSHRA paradigm, we employed non-expert listeners using their own audio equipment—with clear instructions to use headphones in a quiet environment—to achieve a cost-effective yet robust evaluation. Previous studies have shown that online testing using webMUSHRA is a reliable alternative to traditional in-person evaluations, even in informal or non-laboratory settings \cite{rumboldCORRELATIONSOBJECTIVESUBJECTIVE, gusó2022lossfunctionsevaluationmetrics, 7471749, ward2018sisec}.
Each participant rated the estimates during a session lasting approximately 30 minutes and received compensation of €5. Participants were recruited via Prolific and were required to have at least one year of musical experience, such as playing an instrument. For each participant, the presentation of individual estimates was randomly shuffled to mitigate ordering effects and reduce potential bias.

We implemented four quality checks on each listener's data: (1) the difference between the hidden reference and the hidden low-pass filtered anchor had to exceed 10 points (on a 0–100 scale), (2) the reference score had to be at least 90, (3) the standard deviation of all ratings given by a single user had to be at least 20, and (4) the time spent rating all estimates of a single stem had to fall within a reasonable time range (e.g., 20–213 seconds). In our study, the hidden reference is the original, unmodified audio sample provided to listeners, while the hidden anchor is a degraded version of the reference, created by applying a low-pass filter. Specifically, for vocals, drums, and other stems, the anchor was produced using a 3500\,Hz low-pass filter; for bass, a 175\,Hz low-pass filter was used to ensure that the difference would be clearly perceptible to listeners. These criteria were selected to ensure participants were attending to key quality differences and to screen out inattentive responses.

A total of 5,889 stem ratings were collected with the following quality check violation distribution: 2,371 ratings (40\%) had zero violations, 1,778 ratings (30\%) had one violation, 1,284 ratings (22\%) had two violations, 396 ratings (7\%) had three violations, and 60 ratings (1\%) had four violations. We chose to include data for which two or fewer quality checks failed; furthermore, we repeated our experiments with strict quality control (no failed quality checks) and found that the overall findings remained unchanged. After applying our quality control criteria, approximately 8\% of the data was dropped. 

\subsection{Correlation Studies}
In line with \cite{Cartwright2018,Torcoli2021}, we computed Kendall’s $\tau$ for each user’s 0–100 ranking of estimates for a given stem and compared these with the corresponding scores produced by a specific objective metric (e.g., SDR). Kendall’s $\tau$ is simply the proportion of concordant pairs between subjective and objective rankings, thereby providing a robust and averageable measure of rank correlation. By calculating Kendall’s $\tau$ on a per-user, per-stem basis, we effectively mitigate confounding factors such as mix bias and individual listener idiosyncrasies. The Kendall’s $\tau$ values were then averaged across the stem-type classes (Vocals, Drums, Bass, and Other) to obtain an overall measure for each metric.

To further explore the influence of error weighting on the correlation between objective metrics and listener expectations, we performed a grid search over the error components of SI-SDR. In our approach, we varied the weight assigned to the interference error (SI-SIR) while assigning its complement to the artifact error (SI-SAR), ensuring that the sum of weights was always 1. For each weight pair, we computed the energy of the interference and artifact errors, combined these energies geometrically using the assigned weights, and then calculated a re-weighted ratio in decibels as described in \eqref{rwsisdr}. Finally, we computed Kendall’s $\tau$ between these re-weighted ratios and the listener rankings, thereby quantifying how variations in the weighting of error components correlate with human perception.

\begin{equation}\label{rwsisdr}
\text{Reweighted SI-SDR} = 10 \log_{10} \left( \frac{\norm{e_{\text{target}}}^{2}}{e_{\text{interference}}^{w} \cdot e_{\text{artifact}}^{\,1-w}} \right)
\end{equation}

For Fréchet Audio Distance, we computed the distance between each stem estimate and its ground truth. Because smaller distances indicate greater similarity, we inverted the distances before calculating Kendall’s $\tau$ against listener scores.

\section{Results and Discussion}
\subsection{Perceptual Rankings of Separation Systems}
\begin{figure}[t]
  \centering
  \centerline{\includegraphics[width=\columnwidth]{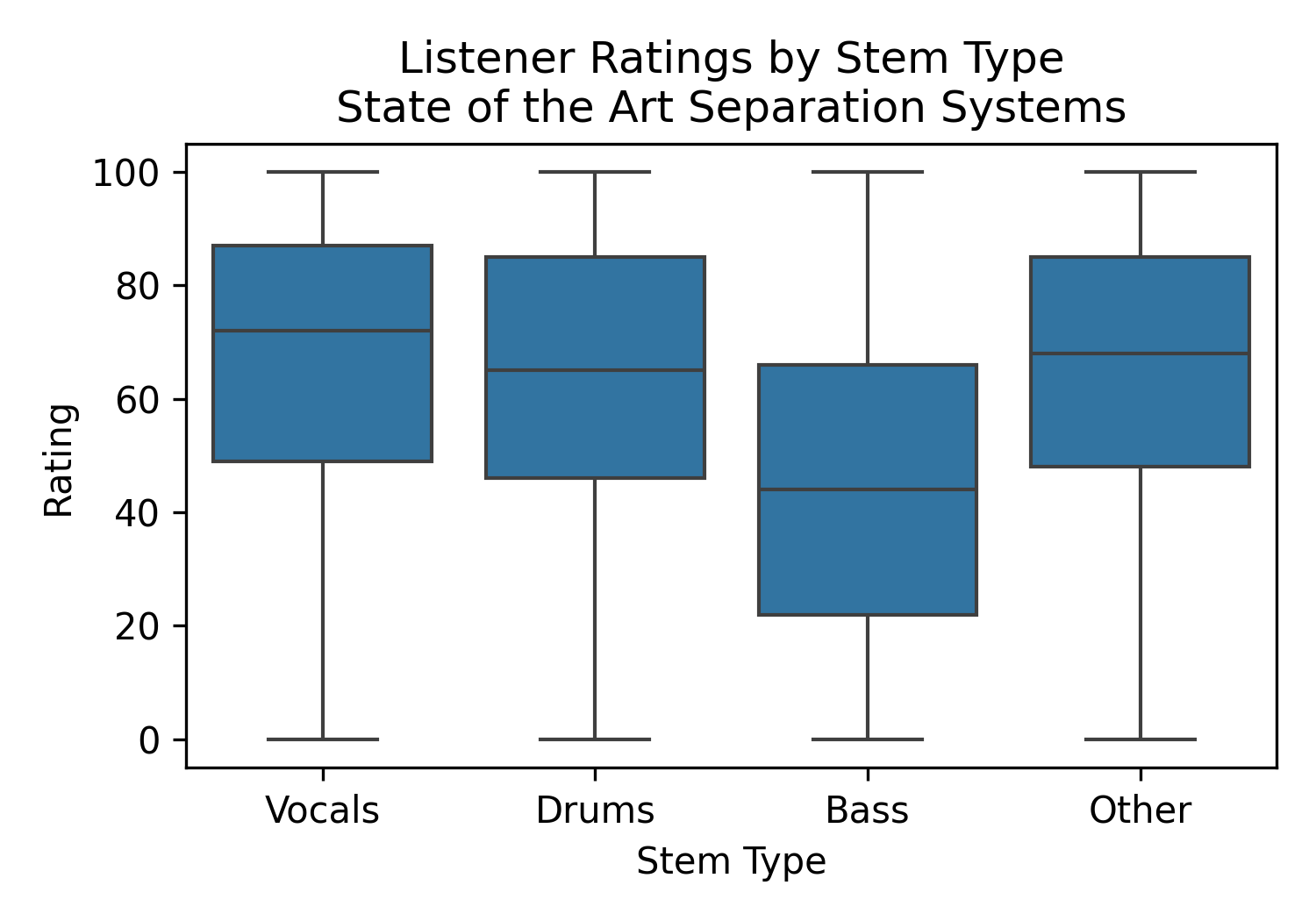}}
\caption{Rating distribution of state-of-the-art systems, by stem type. State-of-the-art models often struggle to produce bass estimates that match the reference, in part because bass instruments, while containing some high-frequency content, are easily masked by other sounds. This makes them especially vulnerable to interference in dense musical textures, leading to noticeably degraded estimates compared to the reference.}
  \label{fig:sota_ratings_by_stem}
\end{figure}
Figure \ref{fig:sota_ratings_by_stem} shows that listeners rate bass estimates produced by state-of-the-art systems (i.e., SCNet-large and HTDemucs-ft) as lower in quality than those for the other stems. We acknowledge that the bass estimates exhibit limited variability compared to other stems—such as vocals or drums—where the best estimates achieve higher scores and the worst are much lower. This suggests that current separation systems are inherently less effective at producing high-quality bass outputs. One contributing factor is that bass instruments, while containing some high-frequency content, are easily masked by other sources in the mix. Additionally, because the bass often functions as part of the rhythm section, its note onsets frequently coincide with broad-spectrum transients from instruments like drums and rhythm guitar, making accurate isolation particularly challenging.

\begin{figure}[t]
  \centering
  \centerline{\includegraphics[width=\columnwidth]{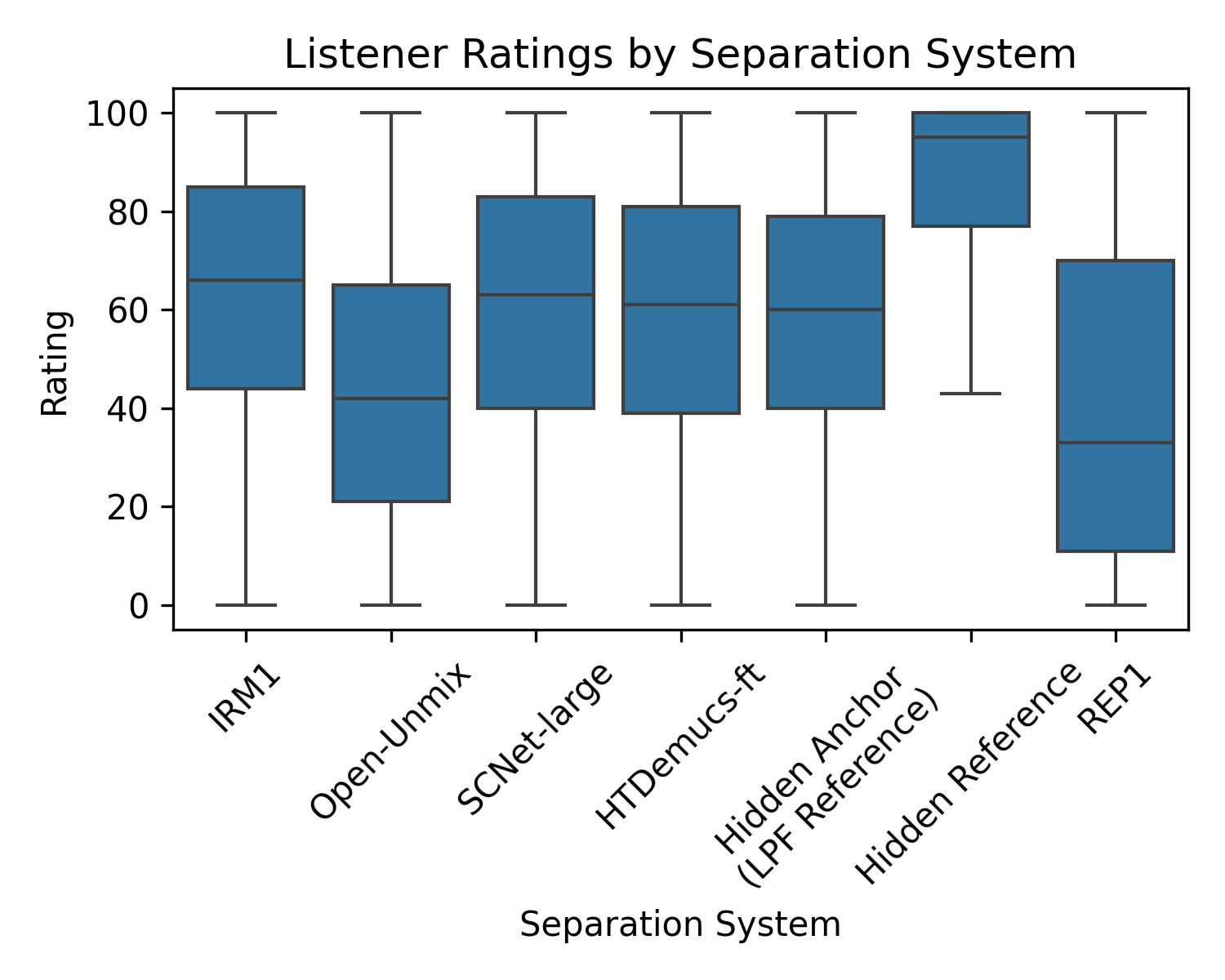}}
\caption{Rating distribution of all stems, by separation system. Listeners clearly distinguish between references, legacy methods such as Open-Unmix and REP1, and state-of-the-art systems such as SCNet-large and HTDemucs-ft; surprisingly, the oracle method IRM1 does not outperform the state-of-the-art systems.}
  \label{fig:separationboxplot}
\end{figure}

Figure \ref{fig:separationboxplot} shows that listeners have a clear preference for estimates produced by state-of-the-art systems, compared to legacy ones. The results confirm that listeners reliably distinguish between separation quality tiers: hidden references are rated highest, followed by state-of-the-art systems (SCNet-large, HTDemucs-ft), then legacy methods (Open-Unmix, REP1). This suggests that even in a web-based setting, participants can meaningfully discriminate separation quality.

Surprisingly, the IRM1 oracle method does not outperform state-of-the-art models. While IRM1 is designed to produce ideal ratio masks based on ground truth data, its outputs may sound overly filtered or musically unnatural—limiting perceived quality. This challenges the assumption that oracle methods reflect an upper bound of perceptual performance, and reinforces the importance of perceptually grounded evaluation methods in music source separation research. IRM1 may benefit from additional post-processing to improve perceptual plausibility, which wasn’t applied here.

\begin{figure}[t]
  \centering
  \centerline{\includegraphics[width=\columnwidth]{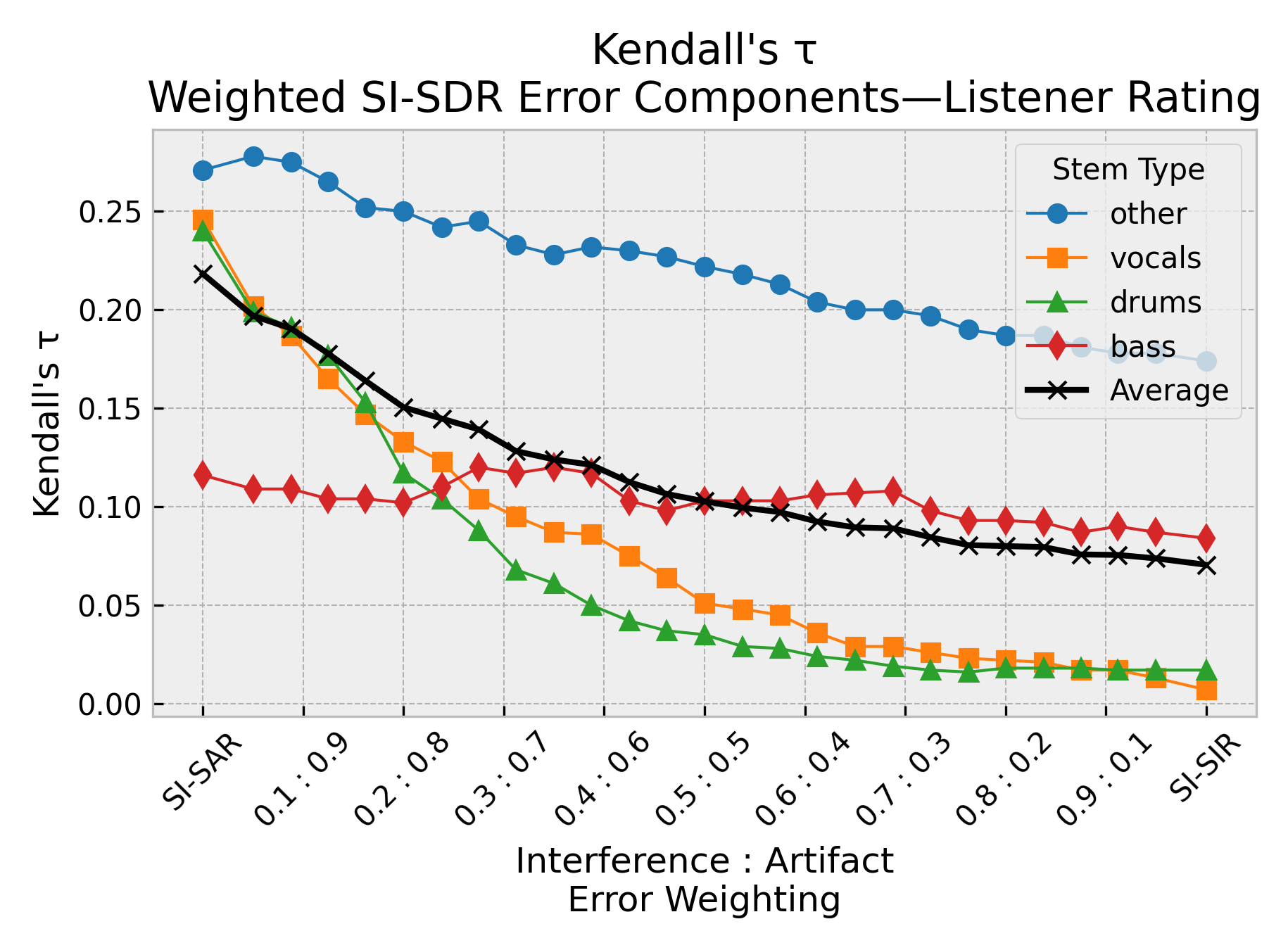}}
\caption{Reweightings of the scale-invariant SDR error terms (SI-SIR vs.\ SI-SAR). Kendall’s $\tau$ decreases almost monotonically as weight shifts from artifact toward interference for all stems except bass, indicating interference errors better predict perception; bass remains flat, likely due to the narrow quality range of bass estimates.}
  \label{fig:gridsearchresult}
\end{figure}

\subsection{Error‐Terms Correlation}
Figure \ref{fig:gridsearchresult} shows how Kendall’s \(\tau\) changes as we sweep the weight from an interference‐only error term (SI-SIR) to an artifact-only error term (SI-SAR) in our reweighted SI-SDR formula (Eq.\,\eqref{rwsisdr}). For Vocals, Drums, and Other stems, \(\tau\) decreases almost monotonically. In contrast, the Bass curve is nearly flat, reflecting the lack of quality variation in our bass estimates. These findings indicate that artifact errors drive listener judgments more strongly than interference errors.

\subsection{Legacy Metrics}
Table \ref{tab:legacymetrics} shows the correlation between BSSEval v4 metrics and human ratings from our listening study. Notably, traditional SDR is the most predictive metric for vocal estimates in all that we tested in this study. We attribute this to the exceptionally high quality of vocal separation in state-of-the-art systems: spatial artifacts and gain mismatches—captured by BSSEval v4’s formulation—become more perceptually salient as the estimates improve. In contrast, scale-invariant (SI-SDR) and scale-dependent SDR (SD-SDR) do not account for spatial error due to their mono-only implementation, likely reducing their effectiveness for evaluating vocals.

While SDR also performs best on average across all stems, some alternative metrics show better alignment for specific sources. For bass, SAR correlates more strongly with listener ratings than any other BSSEval v4 metric, suggesting that perceived artifacts are a primary driver of quality judgments in this stem.

\begin{table}[t]
\centering
\caption{Kendall's $\tau$ for BSSEval v4}
\label{tab:legacymetrics}
\begin{tabularx}{\columnwidth}{l*{4}{>{\centering\arraybackslash}X}}
\toprule
Stem Type & SDR & ISR & SAR & SIR \\
\midrule
Vocals   & \textbf{0.316} & 0.262 & 0.258 & 0.236 \\
Drums    & 0.165 & \textbf{0.169} & 0.124 & 0.067 \\
Bass     & 0.086 & 0.058 & \textbf{0.181} & 0.093 \\
Other    & \textbf{0.273} & 0.176 & 0.199 & 0.213 \\
\midrule
\textbf{Average}  & \textbf{0.211} & 0.167 & 0.190 & 0.152 \\
\bottomrule
\end{tabularx}
\end{table}

\begin{table}[t]
\centering
\caption{Kendall's $\tau$ for SD-SDR, SI-SDR, SI-SAR, and SI-ISR}
\label{tab:si_sdr_metrics}
\begin{tabularx}{\columnwidth}{l*{4}{>{\centering\arraybackslash}X}}
\toprule
Stem Type & SD-SDR & SI-SDR & SI-SAR & SI-ISR \\
\midrule
Vocals   & 0.150 & 0.197 & \textbf{0.246} & 0.007 \\
Drums    & 0.133 & 0.203 & \textbf{0.240} & 0.017 \\
Bass     & 0.080 & 0.084 & \textbf{0.116} & 0.084 \\
Other    & 0.261 & \textbf{0.277} & 0.271 & 0.174 \\
\midrule
\textbf{Average}  & 0.156 & 0.190 & \textbf{0.218} & 0.071 \\
\bottomrule
\end{tabularx}
\end{table}

In contrast, Table \ref{tab:si_sdr_metrics} presents correlations for scale-invariant metrics as implemented in the Northwestern University Source Separation Library (nussl)~\cite{nussl}. While SI-SAR shows improved performance overall—particularly for instruments—it falls short of traditional SDR for vocals, reinforcing the importance of scale and spatial fidelity in that stem. Interestingly, for drums and bass, SI-SAR outperforms both SDR and SAR from BSSEval, indicating its utility for sources where distortion and interference dominate perceptual quality. This is likely because SI-SAR excludes interference and noise from the numerator—unlike traditional SAR—making it a more focused and predictive measure of artifact-related degradation.

Overall, these comparisons underscore that no single metric uniformly captures human perception across all stem types. While traditional SDR remains the most faithful for vocals, modern metrics like SI-SAR offer better alignment for instrument stems—particularly when masking and artifacts dominate perceptual judgments.

\subsection{Fréchet Audio Distance}
\begin{table}[t]
\centering
\caption{Kendall's $\tau$ for Various Embeddings by Stem Type\\
CLAP-LAION-music outperforms all other embeddings across every stem type. However, none of the embeddings yield positive correlation with listener ratings for vocal estimates.}
\label{tab:embedding_metrics}
\begin{tabularx}{\columnwidth}{l*{5}{>{\centering\arraybackslash}X}}
\toprule
Stem Type & CLAP-LAION-music & EnCodec & VGGish & Wave2Vec2 & HuBERT \\
\midrule
Vocals   & 0.000 & -0.092  & -0.129 & -0.096    & -0.118 \\
Drums    & \textbf{0.253} & 0.130   & 0.046  & -0.010    & 0.114  \\
Bass     & \textbf{0.188} & 0.151   & 0.098  & 0.056     & 0.064  \\
Other    & \textbf{0.198} & 0.017   & 0.099  & 0.022     & 0.073  \\
\midrule
\textbf{Average}  & \textbf{0.160} & 0.052   & 0.029  & -0.008    & 0.034  \\
\bottomrule
\end{tabularx}
\end{table}


None of the evaluated embeddings produced a positive Kendall’s $\tau$ for vocal estimates. We initially hypothesized that embeddings trained on speech, such as HuBERT and Wave2Vec2, might better capture degradations in vocal stems—but this was not supported by our results. However, \texttt{CLAP-LAION-music} consistently outperformed others on instrument stems—substantially improving over VGGish, the default embedding in FAD. This aligns with recent work showing that CLAP-based FAD correlates better with perceptual audio quality than its VGGish-based counterpart \cite{PAM, HwangEtAl2025}. For drums, a Kendall’s $\tau$ of 0.25 matches the best-performing energy metric (SI-SAR); for bass, $\tau = 0.19$ is on par with SAR and well above standard SDR. For the other estimates, however, energy-based metrics (SDR, SI-SDR, SI-SAR) yield Kendall’s $\tau$ values around 0.27—slightly outperforming the 0.20 achieved by the \texttt{CLAP-LAION-music} embedding. 

While VGGish captures aggregated mel-spectral textures, CLAP adds high-level semantic content, HuBERT focuses on phonetic patterns, and Wave2Vec2 is tuned to phonetic articulations. However, none of these embeddings necessarily capture errors related to spatial distortion, interference, or artifacts. Systematic ablations—for example, masking embedding dimensions associated with pitch or noise—could help identify the perceptual drivers within the embedding space and guide improvements to FAD for source separation evaluation.

\section{Conclusion}
Our work builds on existing literature by conducting a large-scale listener study on the MUSDB18 dataset, comparing a diverse set of separation systems—including state-of-the-art deep-learning models (e.g., HTDemucs-ft, SCNet-large), an oracle method (IRM1), and legacy methods (REP1, Open-Unmix)—against a range of objective metrics. Our comprehensive evaluation, using a webMUSHRA-based online test with carefully implemented quality checks and rigorous statistical analysis, confirms that traditional metrics such as SDR often fail to capture perceptual quality. In particular, we found that the scale-invariant signal-to-artifacts ratio (SI-SAR) correlates best with human ratings across instrument stems, especially for drums and bass, where it either matches or surpasses legacy metrics.

Furthermore, our grid search over the weighting parameters of SI-SDR revealed that artifact errors are more highly correlated with listener perception than interference errors. Our analysis of Fréchet Audio Distance embeddings shows that, while the CLAP-LAION-music embedding yields promising results for instrument stems—with drum estimates achieving Kendall's $\tau = 0.25$ and bass estimates $\tau = 0.19$, on par with the best bass metric (SAR)—none of the embeddings produced positive correlations for vocal estimates. This highlights that the embeddings used by FAD do not capture the perceptually relevant details of vocal quality, and further refinement of perceptual metrics is necessary.

Overall, our findings underscore the critical importance of using robust, listener-informed evaluations to benchmark objective metrics in music source separation. By releasing our complete raw listener data, we aim to promote transparency and support future meta-analyses that can further improve both evaluation techniques and separation technologies.
All data used in this study are publicly available at \url{https://doi.org/10.5281/zenodo.15843081}


\IEEEtriggeratref{15}
\bibliographystyle{IEEEtran}
\bibliography{WASPAA2025_paper}

\end{document}